\RequirePackage{lineno}
\documentclass[aps,twocolumn,showpacs,byrevtex,prl,reprint]{revtex4-1}

\usepackage{graphicx}
\usepackage{dcolumn}
\usepackage{bm}
\usepackage{rotating}
\usepackage{epstopdf}
\usepackage{color}
\usepackage{verbatim} 
\usepackage{multirow}
\usepackage[abs]{overpic}
\usepackage{amsmath}
\usepackage{amssymb}
\usepackage{xspace}
\usepackage[titletoc]{appendix}
\usepackage{float}

\usepackage[colorlinks,
            linkcolor=blue,
            anchorcolor=blue,
            citecolor=blue]{hyperref}

\newcommand{\PreserveBackslash}[1]{\let\temp=\\#1\let\\=\temp}
\newcolumntype{C}[1]{>{\PreserveBackslash\centering}p{#1}}
\newcolumntype{R}[1]{>{\PreserveBackslash\raggedleft}p{#1}}
\newcolumntype{L}[1]{>{\PreserveBackslash\raggedright}p{#1}}

\newcommand{\EE}{e^+e^-}

\newcommand{\too}{\rightarrow}

\begin{document}
\graphicspath{{figure/}}
\DeclareGraphicsExtensions{.eps,.png,.ps}
\title{\boldmath First study of antihyperon-nucleon scattering $\bar{\Lambda}p\too\bar{\Lambda}p$ and measurement of $\Lambda p\too\Lambda p$ cross section}
\author{
  \begin{small}
    \begin{center}
      M.~Ablikim$^{1}$, M.~N.~Achasov$^{4,c}$, P.~Adlarson$^{75}$, O.~Afedulidis$^{3}$, X.~C.~Ai$^{80}$, R.~Aliberti$^{35}$, A.~Amoroso$^{74A,74C}$, Q.~An$^{71,58,a}$, Y.~Bai$^{57}$, O.~Bakina$^{36}$
% ...
\\
\vspace{0.2cm}
(BESIII Collaboration)\\
\vspace{0.2cm} {\it
$^{1}$ Institute of High Energy Physics, Beijing 100049, People's Republic of China\\
% ...
$^{3}$ Bochum  Ruhr-University, D-44780 Bochum, Germany\\
$^{4}$ Budker Institute of Nuclear Physics SB RAS (BINP), Novosibirsk 630090, Russia\\
% ...
$^{35}$ Johannes Gutenberg University of Mainz, Johann-Joachim-Becher-Weg 45, D-55099 Mainz, Germany\\
$^{36}$ Joint Institute for Nuclear Research, 141980 Dubna, Moscow region, Russia\\
% ...
$^{57}$ Southeast University, Nanjing 211100, People's Republic of China\\
$^{58}$ State Key Laboratory of Particle Detection and Electronics, Beijing 100049, Hefei 230026, People's Republic of China\\
% ...
$^{71}$ University of Science and Technology of China, Hefei 230026, People's Republic of China\\
% ...
$^{74}$ University of Turin and INFN, (A)University of Turin, I-10125, Turin, Italy; (B)University of Eastern Piedmont, I-15121, Alessandria, Italy; (C)INFN, I-10125, Turin, Italy\\
$^{75}$ Uppsala University, Box 516, SE-75120 Uppsala, Sweden\\
% ...
$^{80}$ Zhengzhou University, Zhengzhou 450001, People's Republic of China\\
\vspace{0.2cm}
$^{a}$ Deceased\\
% ...
$^{c}$ Also at the Novosibirsk State University, Novosibirsk, 630090, Russia\\
% ...
      }\end{center}
    \vspace{0.4cm}
\end{small}
}
\affiliation{}


\begin{abstract}
Using $(10.087\pm0.044)\times10^{9}$ $J/\psi$ events collected with the BESIII detector at the BEPCII storage ring, the processes $\Lambda p\too\Lambda p$ and $\bar{\Lambda}p\too\bar{\Lambda}p$ are studied, where the $\Lambda/\bar{\Lambda}$ baryons are produced in the process $J/\psi\too\Lambda\bar{\Lambda}$ and the protons are the hydrogen nuclei in the cooling oil of the beam pipe. Clear signals are observed for the two reactions. The cross sections in $-0.9\leq\rm{cos}\theta_{\Lambda/\bar{\Lambda}}\leq0.9$ are measured to be $\sigma(\Lambda p\too\Lambda p)=(12.2\pm1.6_{\rm{stat}}\pm1.1_{\rm{sys}})$~mb and $\sigma(\bar{\Lambda} p\too\bar{\Lambda} p)=(17.5\pm2.1_{\rm{stat}}\pm1.6_{\rm{sys}})$~mb at the $\Lambda/\bar{\Lambda}$ momentum of $1.074$~GeV/$c$ within a range of $\pm0.017$~GeV/$c$, where the $\theta_{\Lambda/\bar{\Lambda}}$ are the scattering angles of the $\Lambda/\bar{\Lambda}$ in the $\Lambda p/\bar{\Lambda}p$ rest frames. Furthermore, the differential cross sections of the two reactions are also measured, where there is a slight tendency of forward scattering for $\Lambda p\too\Lambda p$, and a strong forward peak for $\bar{\Lambda}p\too\bar{\Lambda}p$. We present an approach to extract the total elastic cross sections by extrapolation. The study of $\bar{\Lambda}p\too\bar{\Lambda}p$ represents the first study of antihyperon-nucleon scattering, and these new measurements will serve as important inputs for the theoretical understanding of the (anti)hyperon-nucleon interaction.
\end{abstract}

\maketitle

One of the main goals of nuclear physics is to understand baryon-baryon interaction in a unified perspective. To achieve this purpose, plentiful nucleon-nucleon (NN) and antinucleon-nucleon ($\rm{\bar{N}}$N) scattering data have been measured~\cite{pdg}. Therefore, the relevant theory of NN and $\rm{\bar{N}}$N interactions is well established, and it can be tightly constrained by experimental data. However, the understanding of hyperon-nucleon (YN) interaction has a large uncertainty due to the lack of relevant measurements. The YN interaction is studied mainly via three methods. The first is to extract the YN correlation functions in heavy-ion collisions~\cite{star1, star2, alice1, alice2}, the second is to study hypernuclei~\cite{hypernuclei1, hypernuclei2, hypernuclei3, hypernuclei4}, and the third is to investigate YN scattering~\cite{scattering1, scattering2, scattering3}. The last method is the most direct way to study YN interaction, but it is limited by the availability and short-lifetime of hyperon beams, leading to a scarcity of YN scattering data~\cite{pdg}. The study of YN interaction is also crucial to determine the equation of state (EoS) of nuclear matter at supersaturation densities and understand the so-called ``hyperon puzzle" of neutron stars (NS)~\cite{neutronstar1, neutronstar2, neutronstar3, neutronstar4, theory1, theory2}. To solve these issues, more YN scattering data is desired to constrain the calculations of YN interaction.

Compared to the YN scattering, the situation is even worse for antihyperon-nucleon ($\rm{\bar{Y}}$N) scattering. Until now, no $\rm{\bar{Y}}$N scattering data have been obtained due to the absence effective antihyperon sources~\cite{pdg}, which results in the very limited related theoretical research. Therefore, the realization of $\rm{\bar{Y}}$N scattering measurements can fill this gap, and new measurements will motivate more effort for the understanding of the $\rm{\bar{Y}}$N interaction. More importantly, $\rm{\bar{Y}}$N scattering data can further constrain the YN interaction theory from another angle.

In this Letter, we present a study of the reactions $\Lambda p\too\Lambda p$ and $\bar{\Lambda}p\too\bar{\Lambda}p$, where $\Lambda$ and $\bar{\Lambda}$ are reconstructed via the decays $\Lambda\too p\pi^-$ and $\bar{\Lambda}\too\bar{p}\pi^+$. The cross sections and differential cross sections of the two reactions are all measured. This is the first study of $\rm{\bar{Y}}$N scattering.

\begin{figure}[htbp]
\begin{center}
\begin{overpic}[width=0.44\textwidth]{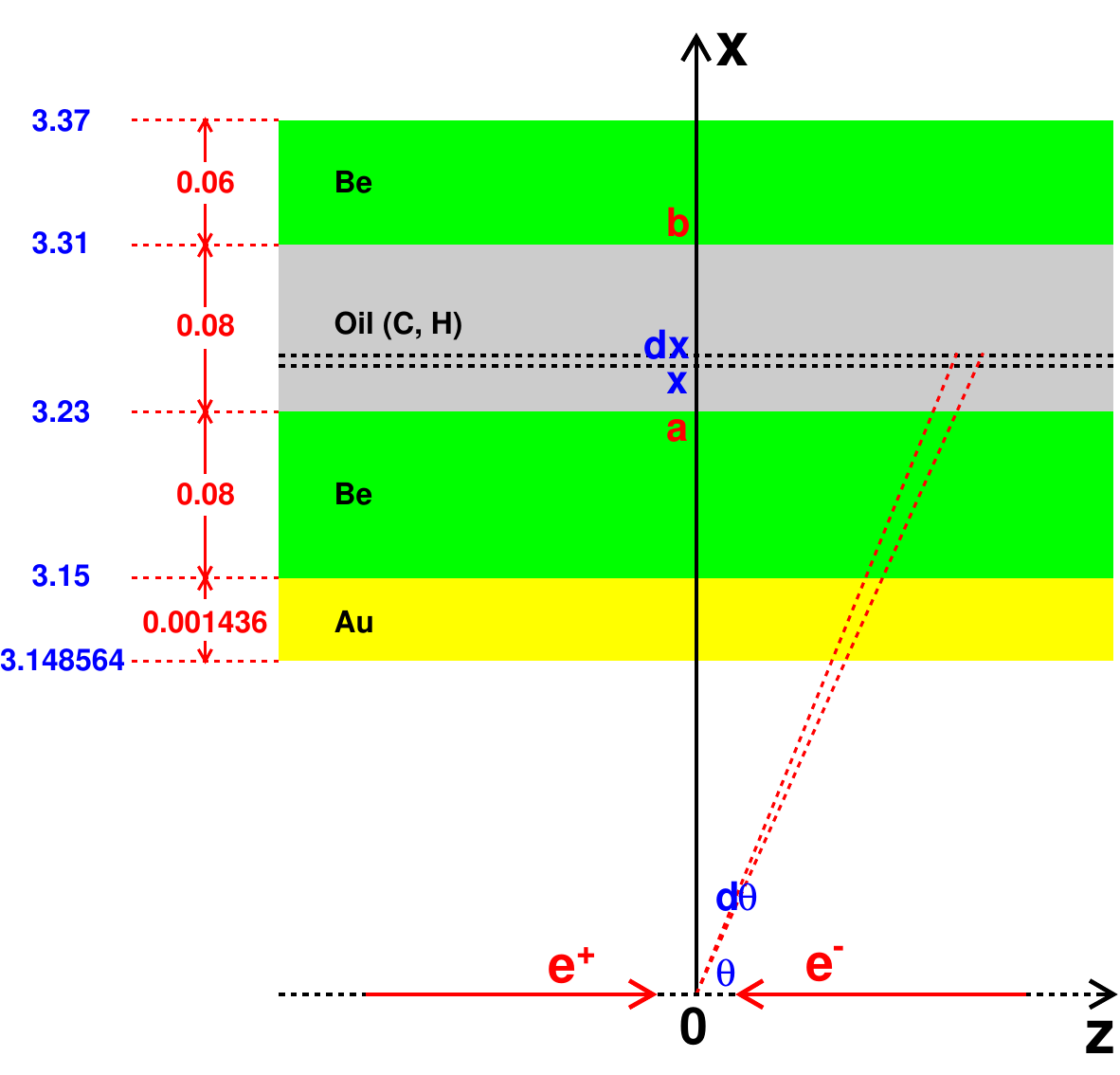}
\end{overpic}
\caption{Schematic diagram of the beam pipe, the length units are centimeter (cm). The $z$-axis is the symmetry axis of the MDC, and the $x$-axis is perpendicular to the $e^+e^-$ beam direction.}
\label{fig:beampipe}
\end{center}
\end{figure}

The BESIII detector records symmetric $e^+e^-$ collisions at the BEPCII collider~\cite{bepcii}. Details of the BESIII detector can be found in Ref.~\cite{besiii}. The material of the beam pipe is composed of gold ($^{197}\rm{Au}$), beryllium ($^{9}\rm{Be}$) and oil $(^{12}\rm{C}:$$^{1}\rm{H}$$=1:2.13)$, as shown in Fig.~\ref{fig:beampipe}. With a sample of $(10.087\pm0.044)\times10^{9}$ $J/\psi$ events collected by the BESIII detector~\cite{totalnumber}, intense almost monoenergetic $\Lambda/\bar{\Lambda}$ hyperons with a momentum of $1.074$~GeV/$c$ within a range of $\pm0.017$~GeV/$c$ can be produced via the decay $J/\psi\too\Lambda\bar{\Lambda}$, the momentum spread is due to the small horizontal crossing angle of $\pm11$~mrad for $e^{\pm}$ beams. Afterwards the $\Lambda/\bar{\Lambda}$ baryons can interact with the material in the beam pipe. A similar idea was proposed forty years ago using $\bar{p}p$ collisions at a LEAR experiment~\cite{lear}. Especially, Ref.~\cite{xiscatteringbes} has used this method to perform the first study of YN interaction using $\Xi^0$-nucleus scattering at BESIII, and $\Lambda$-nucleus scattering was measured in Ref.~\cite{lambdascatteringbes}. Furthermore, utilizing the almost static protons in the $^{1}\rm{H}$ of the cooling oil of the beam pipe, the information on the interaction between (anti)hyperon and proton can be directly extracted via (anti)hyperon-proton scattering in this way.

In this analysis, simulated data samples are produced with a {\sc{Geant4}}-based~\cite{geant4} Monte Carlo (MC) package, which includes the geometric description of the BESIII detector~\cite{besiii} and the detector response. They are used to determine detection efficiencies and to estimate backgrounds. The simulation models the beam energy spread and initial state radiation (ISR) in the $e^+e^-$ annihilations with the generator {\sc kkmc}~\cite{KKMC}. The inclusive MC sample includes both the production of the $J/\psi$ resonance and the continuum processes incorporated in {\sc kkmc}~\cite{KKMC}. All particle decays are modeled with {\sc evtgen}~\cite{ref:evtgen} using branching fractions either taken from the Particle Data Group (PDG)~\cite{pdg}, where available, or otherwise estimated with {\sc lundcharm}~\cite{ref:lundcharm}. Final state radiation (FSR) from charged final state particles is incorporated using the {\sc photos} package~\cite{photos}. The signal process considered in this analysis is $J/\psi\too\Lambda\bar{\Lambda}$ with either $\Lambda p\too\Lambda p$ or $\bar{\Lambda}p\too\bar{\Lambda}p$, $\Lambda\too p\pi^-$, $\bar{\Lambda}\too \bar{p}\pi^+$. In the signal simulation, the angular distribution of $J/\psi\too\Lambda\bar{\Lambda}$ is generated according to the measurement in Ref.~\cite{alpha}. We simulate the reactions $\Lambda p\too\Lambda p/\bar{\Lambda}p\too\bar{\Lambda}p$ by taking the proton to be at rest, and the hyperon angular distribution is generated using an isotropic phase-space distribution to obtain the angle dependent detection efficiency.

Charged tracks detected in the multilayer drift chamber (MDC) are required to be within a polar angle ($\theta$) range of $|\cos\theta|<0.93$, where $\theta$ is the angle between the charged track and the $z$-axis, which is the symmetry axis of the MDC. Particle identification for charged tracks combines measurements of the energy loss (d$E$/d$x$) in the MDC and the flight time in the time-of-flight system (TOF) to form likelihoods $\mathcal{L}(h)$ $(h=p, K, \pi)$ for each hadron $h$ hypothesis. Tracks are identified as protons when the proton hypothesis has the greatest likelihood $(\mathcal{L}(p)>\mathcal{L}(\pi)$ and $\mathcal{L}(p)>\mathcal{L}(K))$, while charged pions are identified by comparing the likelihoods for the pion hypotheses, $(\mathcal{L}(\pi)>\mathcal{L}(K)$ and $\mathcal{L}(\pi)>\mathcal{L}(p))$.

Since the final states of the two reactions all contain $pp\bar{p}\pi^{+}\pi^{-}$, candidate events must have five charged tracks, and two $p$, one $\bar{p}$, one $\pi^+$ and one $\pi^-$ are required to be identified. For the decay $\bar{\Lambda}\too \bar{p}\pi^+$, we perform a vertex fit to the $\bar{p}\pi^+$ combination, and the $\bar{\Lambda}$ signal region is defined as $|M(\bar{p}\pi^+)-m_{\bar{\Lambda}}|<0.003$~GeV/$c^{2}$, where $m_{\bar{\Lambda}}$ is the nominal mass of the $\bar{\Lambda}$. In this Letter, all nominal masses are taken from PDG~\cite{pdg}. For the decay $\Lambda\too p\pi^-$, we perform the vertex fit by considering both $p\pi^-$ combinations. The $p\pi^-$ combination with the smallest value of $|M(p\pi^-)-m_{\Lambda}|$, where $m_{\Lambda}$ is the $\Lambda$ nominal mass, is taken as $\Lambda$ candidate. The $\Lambda$ signal region is defined as $|M(p\pi^-)-m_{\Lambda}|<0.003$~GeV/$c^{2}$. Finally, a vertex fit is performed to the combination of the $\Lambda/\bar{\Lambda}$ and the remaining $p$ for the reactions $\Lambda p\too\Lambda p/\bar{\Lambda}p\too\bar{\Lambda}p$.

To select the signal events of $J/\psi\too\Lambda\bar{\Lambda}$, the invariant mass recoiling against the $\bar{\Lambda}/\Lambda$, $M_{\text{recoil}}(\bar{\Lambda}/\Lambda)$, is required to be in the $\Lambda/\bar{\Lambda}$ signal region, defined as $[m_{\Lambda/\bar{\Lambda}}-0.020, m_{\Lambda/\bar{\Lambda}}+0.016]$~GeV/$c^{2}$, where $M_{\text{recoil}}(\bar{\Lambda}/\Lambda)\equiv\sqrt{E^2_{\text{beam}}-|\vec{p}_{\bar{\Lambda}/\Lambda}c|^2}/c^2$, $E_{\text{beam}}$ is the $e^{\pm}$ beam energy, and $\vec{p}_{\bar{\Lambda}/\Lambda}$ is the measured momentum of the $\bar{\Lambda}/\Lambda$ candidate in the $\EE$ rest frame. The main background is $J/\psi\too\Lambda\bar{\Lambda}$, $\Lambda\too p\pi^-$, $\bar{\Lambda}\too\bar{p}\pi^+$, where no scattering of $\Lambda/\bar{\Lambda}$ with a proton from the beam pipe occured. To suppress this background, the recoil mass of $\bar{\Lambda}p_{\Lambda}/\Lambda\bar{p}$, $M_{\text{recoil}}(\bar{\Lambda}p_{\Lambda}/\Lambda\bar{p})$, is obtained from the four-momenta of the initial $e^+e^-$ system and the $\bar{\Lambda}/\Lambda$ and $p_{\Lambda}/\bar{p}$ candidates, where $p_{\Lambda}$ is the proton from $\Lambda$ decays. $M_{\text{recoil}}(\bar{\Lambda}p_{\Lambda}/\Lambda\bar{p})$ should be around the nominal $\pi^-/\pi^+$ mass for this background, so we require $M_{\text{recoil}}(\bar{\Lambda}p_{\Lambda}/\Lambda\bar{p})<0$~GeV/$c^2$ to remove these events. To select those signal events that react with the cooling oil in the beam pipe, the $R_{xy}$ signal region is defined as $[3.0, 3.5]$~cm, taking into account the detector resolution, where $R_{xy}$ is the distance from the reconstructed $\Lambda p/\bar{\Lambda}p$ vertex to the $z$-axis. To remove the events from the reactions between $\Lambda/\bar{\Lambda}$ and $^{197}\rm{Au}/^{9}\rm{Be}/^{12}\rm{C}$ nuclei, we define the momentum of the proton in the $^{1}\rm{H}$ of the cooling oil as $P(p_{\rm{oil}})\equiv|\vec{P}_{\Lambda/\bar{\Lambda}}+\vec{P}_{p}-(\vec{P}_{e^+e^-}-\vec{P}_{\bar{\Lambda}/\Lambda})|$, where $\vec{P}$ represents the momentum of each particle in the lab frame. Because the proton in the $^{1}\rm{H}$ of the cooling oil is practically static, while the proton in the $^{197}\rm{Au}/^{9}\rm{Be}/^{12}\rm{C}$ nuclei has Fermi momentum, the $P(p_{\rm{oil}})$ should be around zero for signal processes but hundreds of MeV/$c$ for background processes. To remove these events, the requirement $P(p_{\rm{oil}})<0.04$~GeV/$c$ is applied.

\begin{figure}[htbp]
\begin{center}
\begin{overpic}[width=0.44\textwidth]{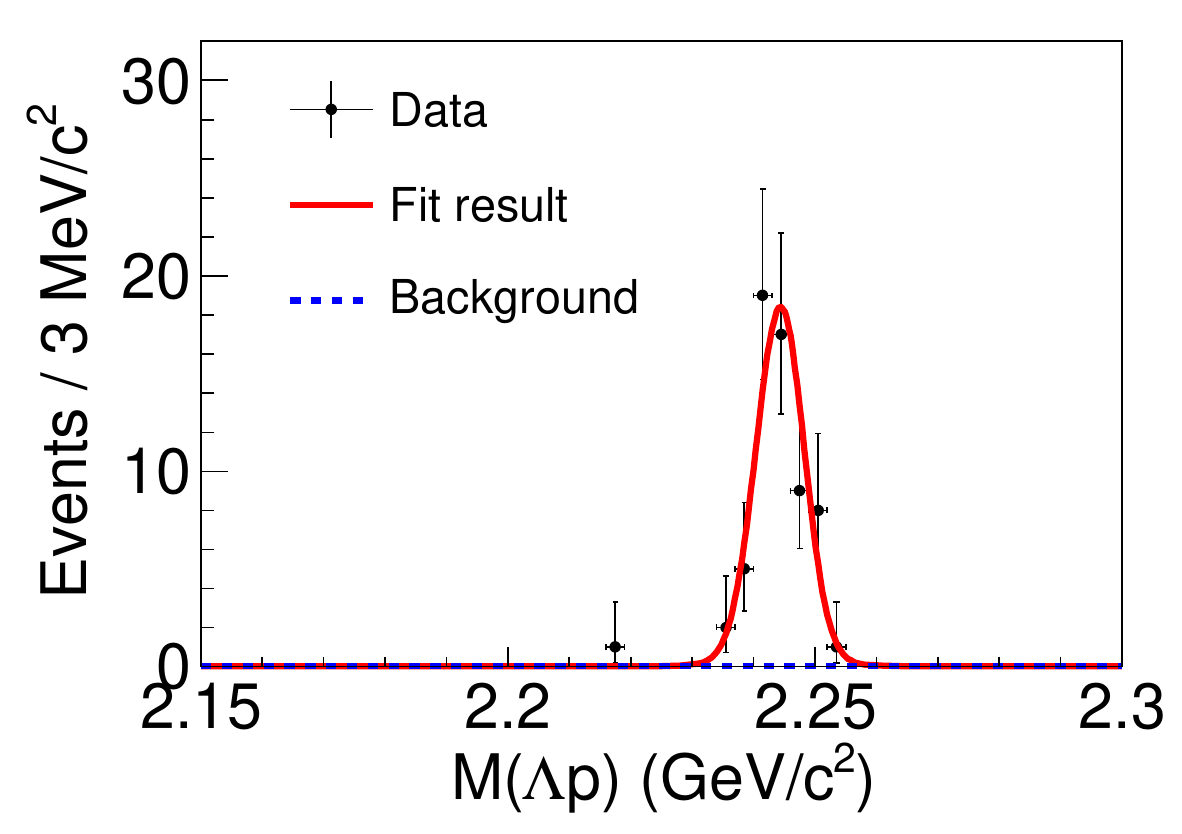}
\end{overpic}
\begin{overpic}[width=0.44\textwidth]{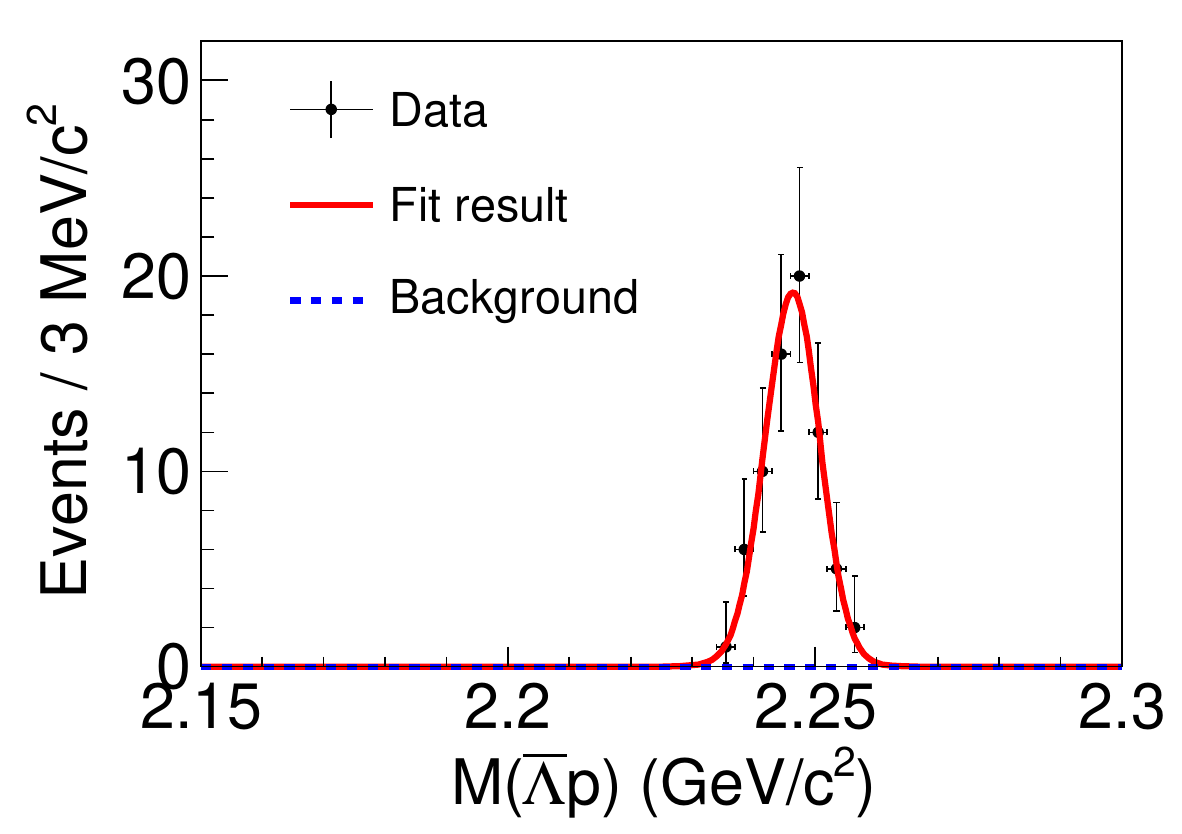}
\end{overpic}
\caption{Distributions of $M(\Lambda p)$ (top) and $M(\bar{\Lambda}p)$ (bottom) of data (black dots with error bars) for the reactions $\Lambda p\too\Lambda p$ and $\bar{\Lambda}p\too\bar{\Lambda}p$, respectively. The red solid curve is the total fit result and the blue dashed curve is the background component.}
\label{fig:fit_result}
\end{center}
\end{figure}

For the signal reactions $\Lambda p\too\Lambda p$ and $\bar{\Lambda}p\too\bar{\Lambda}p$ produced from the decay $J/\psi\too\Lambda\bar{\Lambda}$, the center-of-mass energies for the incident $\Lambda/\bar{\Lambda}$ and a static $p$ are all $2.243$~GeV/$c^2$ within a range of $\pm0.005$~GeV/$c^2$. Figure~\ref{fig:fit_result} shows the $M(\Lambda p)$ and $M(\bar{\Lambda}p)$ distributions from data after the final event selection. Clear enhancements are seen around $2.243$~GeV/$c^2$, corresponding to the reactions $\Lambda p\too\Lambda p$ and $\bar{\Lambda}p\too\bar{\Lambda}p$, respectively. A detailed study of the $J/\psi$ inclusive MC sample shows that there is no peaking background in the signal region. To determine the signal yield, an unbinned maximum likelihood fit is performed to the $M(\Lambda p)$ distribution and $M(\bar{\Lambda}p)$ distribution, respectively. We use the MC-determined shape convolved with a free Gaussian function to describe the signal, where the yield acts as a free fit parameter. The free Gaussian function is used to describe the difference in the data and signal MC resolutions. The background is described by a uniform distribution with the number of events as free parameter. The fit results are shown in Fig.~\ref{fig:fit_result}. The signal yields returned by the fits are $N^{\rm{sig}}_{\Lambda p}=60.9\pm7.8$ and $N^{\rm{sig}}_{\bar{\Lambda}p}=72.0\pm8.5$ for the reactions $\Lambda p\too\Lambda p$ and $\bar{\Lambda}p\too\bar{\Lambda}p$, respectively, and the goodness of the fits for the two reactions are $\chi^2/$ndf = 4.8/4 = 1.2 and 0.8/4 = 0.2 without considering empty bins.

To extract the differential cross sections for the two reactions, we need the signal yields as a function of cos$\theta_{\Lambda/\bar{\Lambda}}$, where $\theta_{\Lambda/\bar{\Lambda}}$ is the scattering angle of the scattered $\Lambda/\bar{\Lambda}$ in the $\Lambda p/\bar{\Lambda}p$ rest frames with the $z$-axis defined by the incident $\Lambda/\bar{\Lambda}$ momentum. Because the efficiency is very low and it is hard to obtain accurate experimental information near the regions cos$\theta_{\Lambda/\bar{\Lambda}}=\pm1$ due to the low momentum of scattered $\Lambda/\bar{\Lambda}$ or $p$, the measurements are restricted to $-0.9\leq\rm{cos}\theta_{\Lambda/\bar{\Lambda}}\leq0.9$. To obtain the number of signal events, we perform a simultaneous fit to the $M(\Lambda p)$ and $M(\bar{\Lambda}p)$ distributions in nine different cos$\theta_{\Lambda/\bar{\Lambda}}$ regions, where the signal shape and background shape are the same as mentioned above. The obtained number of signal events in the nine cos$\theta_{\Lambda/\bar{\Lambda}}$ regions are summarized in Table~\ref{tab:result}. It is worth mentioning that no events survived in the $-1.0<\rm{cos}\theta_{\Lambda/\bar{\Lambda}}<-0.9$ and the $0.9<\rm{cos}\theta_{\Lambda/\bar{\Lambda}}<1.0$ regions for data.
\begin{table}[htbp]
\renewcommand{\arraystretch}{1.25}
\footnotesize
\begin{center}
\caption{Relevant parameters for the differential cross sections, where cos$\theta_{\Lambda/\bar{\Lambda}}$ is the scattering angle, $N^{\rm{sig}}_i$ is the number of signal events, $\epsilon_i$ is the efficiency, $\frac{d\sigma}{d\Omega}$ is the differential cross section, and $i$ represents the different cos$\theta_{\Lambda/\bar{\Lambda}}$ bins. The first value in parentheses is for $\Lambda p\too\Lambda p$, and the second for $\bar{\Lambda}p\too\bar{\Lambda}p$.}
\label{tab:result}
\begin{tabular}{cccc}
  \hline
  \hline
  cos$\theta_{\Lambda/\bar{\Lambda}}$ & $N^{\rm{sig}}_i$ & $\epsilon_i$ (\%) & $\frac{d\sigma}{d\Omega}$ (mb/sr) \\
  \hline
  $[-0.9, -0.7]$ & $(5.0^{+2.6}_{-1.9}, 0.0^{+1.1}_{-0.0})$   & $(6.94, 4.93)$   & $(1.7^{+0.9}_{-0.7}, 0.0^{+0.5}_{-0.0})$ \\
  $(-0.7, -0.5]$ & $(1.0^{+1.4}_{-0.7}, 0.0^{+1.1}_{-0.0})$   & $(14.13, 10.44)$ & $(0.2^{+0.2}_{-0.1}, 0.0^{+0.3}_{-0.0})$ \\
  $(-0.5, -0.3]$ & $(1.0^{+1.4}_{-0.7}, 1.0^{+1.4}_{-0.7})$   & $(17.32, 13.27)$ & $(0.2^{+0.2}_{-0.1}, 0.2^{+0.3}_{-0.1})$ \\
  $(-0.3, -0.1]$ & $(11.0^{+3.7}_{-3.0}, 0.0^{+1.1}_{-0.0})$  & $(17.74, 14.66)$ & $(1.5^{+0.5}_{-0.4}, 0.0^{+0.2}_{-0.0})$ \\
  $(-0.1, 0.1]$  & $(6.9^{+3.0}_{-2.3}, 0.0^{+1.1}_{-0.0})$   & $(19.11, 15.79)$ & $(0.9^{+0.4}_{-0.3}, 0.0^{+0.2}_{-0.0})$ \\
  $(0.1, 0.3]$   & $(5.0^{+2.6}_{-1.9}, 2.0^{+1.8}_{-1.1})$   & $(19.53, 16.82)$ & $(0.6^{+0.3}_{-0.2}, 0.3^{+0.3}_{-0.2})$ \\
  $(0.3, 0.5]$   & $(12.0^{+3.8}_{-3.1}, 7.0^{+3.0}_{-2.3})$  & $(19.21, 17.68)$ & $(1.5^{+0.5}_{-0.4}, 1.0^{+0.4}_{-0.3})$ \\
  $(0.5, 0.7]$   & $(13.0^{+3.9}_{-3.3}, 25.0^{+5.3}_{-4.7})$ & $(19.71, 17.60)$ & $(1.6^{+0.5}_{-0.4}, 3.4^{+0.7}_{-0.6})$ \\
  $(0.7, 0.9]$   & $(6.0^{+2.8}_{-2.1}, 37.0^{+6.4}_{-5.8})$  & $(9.80, 9.93)$   & $(1.5^{+0.7}_{-0.5}, 9.0^{+1.6}_{-1.4})$ \\
  \hline
  \hline
\end{tabular}
\end{center}
\end{table}

Using the same method as in Ref.~\cite{xiscatteringbes}, the cross sections of the reactions $\Lambda p\too\Lambda p$ and $\bar{\Lambda}p\too\bar{\Lambda}p$ can be determined, the only difference is that we use the proton in the $^{1}\rm{H}$ of the cooling oil of the beam pipe as the target material. The total elastic cross sections are calculated with
\begin{equation}
    \sigma(\Lambda p\too\Lambda p/\bar{\Lambda}p\too\bar{\Lambda}p) = \frac{N^{\rm{sig}}_{\Lambda p/\bar{\Lambda}p}} {\epsilon_{\Lambda p/\bar{\Lambda}p} \mathcal{B} \mathcal{L}_{\rm{eff}} },
\end{equation}
where $\epsilon_{\Lambda p/\bar{\Lambda}p}=\frac{\sum\limits_{i}\epsilon_{i}(\frac{d\sigma}{d\Omega})_{i}}{\sum\limits_{i}(\frac{d\sigma}{d\Omega})_{i}}$ is the weighted selection efficiency according to the differential cross section distribution, which will be introduced later. $\mathcal{B}$ is the product of the branching ratios of the intermediate states, defined as $\mathcal{B}\equiv\mathcal{B}(\Lambda\too p\pi^-)\mathcal{B}(\bar{\Lambda}\too\bar{p}\pi^+)$, and $\mathcal{L}_{\rm{eff}}$ is the effective luminosity of the reaction of the $\Lambda/\bar{\Lambda}$ flux produced from $J/\psi\too\Lambda\bar{\Lambda}$ with the target material:
\begin{equation}
    \mathcal{L}_{\rm{eff}} = \frac{\it{N}_{\it{J}/\psi}\mathcal{B}_{\it{J}/\psi}}{\rm{2}+\frac{\rm{2}}{\rm{3}}\alpha} \int_{a}^{b}\int_{\rm{0}}^{\pi} (\rm{1}+\alpha \rm{cos}^2\theta) \it{e}^{-\frac{x}{\rm{sin}\theta \it{\beta\gamma L}}}N_{H}\rm{d}\theta \rm{d}\it{x}.
\end{equation}
In the integral of this formula, the angular distribution of the $\Lambda/\bar{\Lambda}$ flux, the attenuation of the $\Lambda/\bar{\Lambda}$ flux and the number of target nuclei are considered. $N_{J/\psi}$ is the number of $J/\psi$ events, $\mathcal{B}_{J/\psi}$ is the branching fraction of $J/\psi\too\Lambda\bar{\Lambda}$, and $\alpha$ is the parameter of the angular distribution of $J/\psi\too\Lambda\bar{\Lambda}$, $\beta\gamma\equiv \frac{\sqrt{E_{\rm{beam}}^{\rm{2}}-m_{\rm{\Lambda/\bar{\Lambda}}}^{\rm{2}}c^{\rm{4}}}}{m_{\rm{\Lambda/\bar{\Lambda}}}c^{\rm{2}}}$ is the ratio of the momentum to the mass of the $\Lambda/\bar{\Lambda}$, and $L\equiv c\tau$ is the product of the speed of light and the mean lifetime of the $\Lambda/\bar{\Lambda}$~\cite{pdg}. $N_{H}$ is the number of target nuclei per unit volume, $a$ and $b$ are the distances from the inner surface and outer surface of the cooling oil in the beam pipe to the $z$-axis, $\theta$ and $x$ are the angle and distance to the $z$-axis, as shown in Fig.~\ref{fig:beampipe}. The beam pipe can be regarded as infinitely long with respect to the product of $\beta\gamma L$ for $\Lambda/\bar{\Lambda}$. The parameters are listed in Table~\ref{tab:result2}, and the corresponding total elastic cross sections in $-0.9\leq\rm{cos}\theta_{\Lambda/\bar{\Lambda}}\leq0.9$ are measured to be $\sigma(\Lambda p\too\Lambda p)=(12.2\pm1.6_{\rm{stat}}\pm1.1_{\rm{sys}})$~mb and $\sigma(\bar{\Lambda} p\too\bar{\Lambda} p)=(17.5\pm2.1_{\rm{stat}}\pm1.6_{\rm{sys}})$~mb at a $\Lambda/\bar{\Lambda}$ momentum of $1.074$~GeV/$c$ within a range of $\pm0.017$~GeV/$c$.
\begin{table}[htbp]
\begin{center}
\caption{Input parameters for the cross section calculations. The first value in brackets is for $\Lambda p\too\Lambda p$, and the second is for $\bar{\Lambda}p\too\bar{\Lambda}p$.}
\label{tab:result2}
\begin{tabular}{cc}
  \hline
  \hline
  Parameter & Result \\
  \hline
  $N^{\rm{sig}}_{\Lambda p/\bar{\Lambda}p}$ & $(60.9\pm7.8, 72.0\pm8.5)$ \\
  $\epsilon_{\Lambda p/\bar{\Lambda}p}$     & $(15.29\%, 12.55\%)$ \\
  $\mathcal{B}$                             & $(40.8321\pm0.4518)\%$~\cite{pdg} \\
  $N_{J/\psi}$                              & $(10.087\pm0.044)\times10^{9}$~\cite{totalnumber} \\
  $\mathcal{B}_{J/\psi}$                    & $(0.189\pm0.009)\%$~\cite{pdg} \\
  $\alpha$                                  & $0.475\pm0.004$~\cite{alpha} \\
  $L$                                       & $(7.89\pm0.06)$~cm~\cite{pdg} \\
  $E_{\rm{beam}}$                           & $1.5485$~GeV \\
  $m_{\Lambda/\bar{\Lambda}}$               & $(1.115683\pm0.000006)$~GeV/$c^2$~\cite{pdg}\\
  $a$                                       & $3.23$~cm~\cite{besiii} \\
  $b$                                       & $3.31$~cm~\cite{besiii} \\
  $N_{H}$                                   & $7.35\times10^{22}~\rm{cm^{-3}}$ \\
  \hline
  \hline
\end{tabular}
\end{center}
\end{table}

\begin{figure}[htbp]
\begin{center}
\begin{overpic}[width=0.44\textwidth]{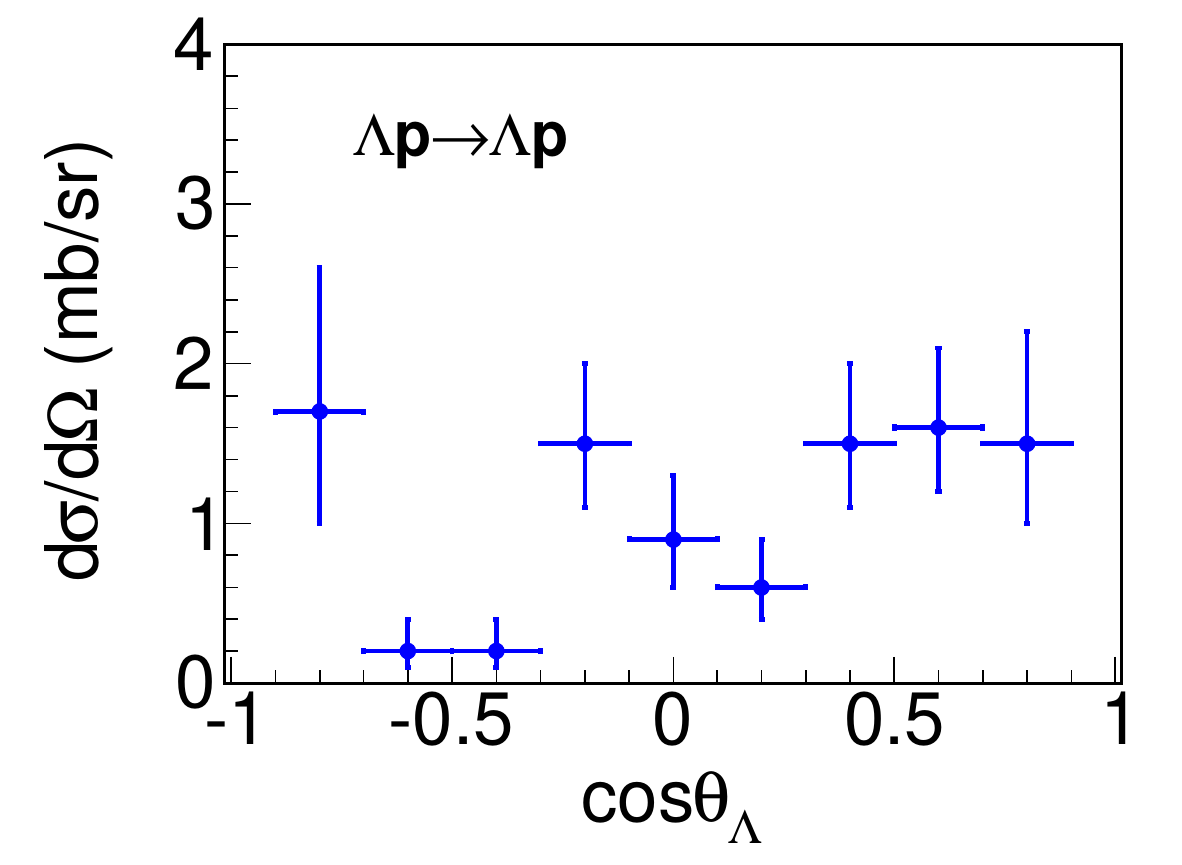}
\end{overpic}
\begin{overpic}[width=0.44\textwidth]{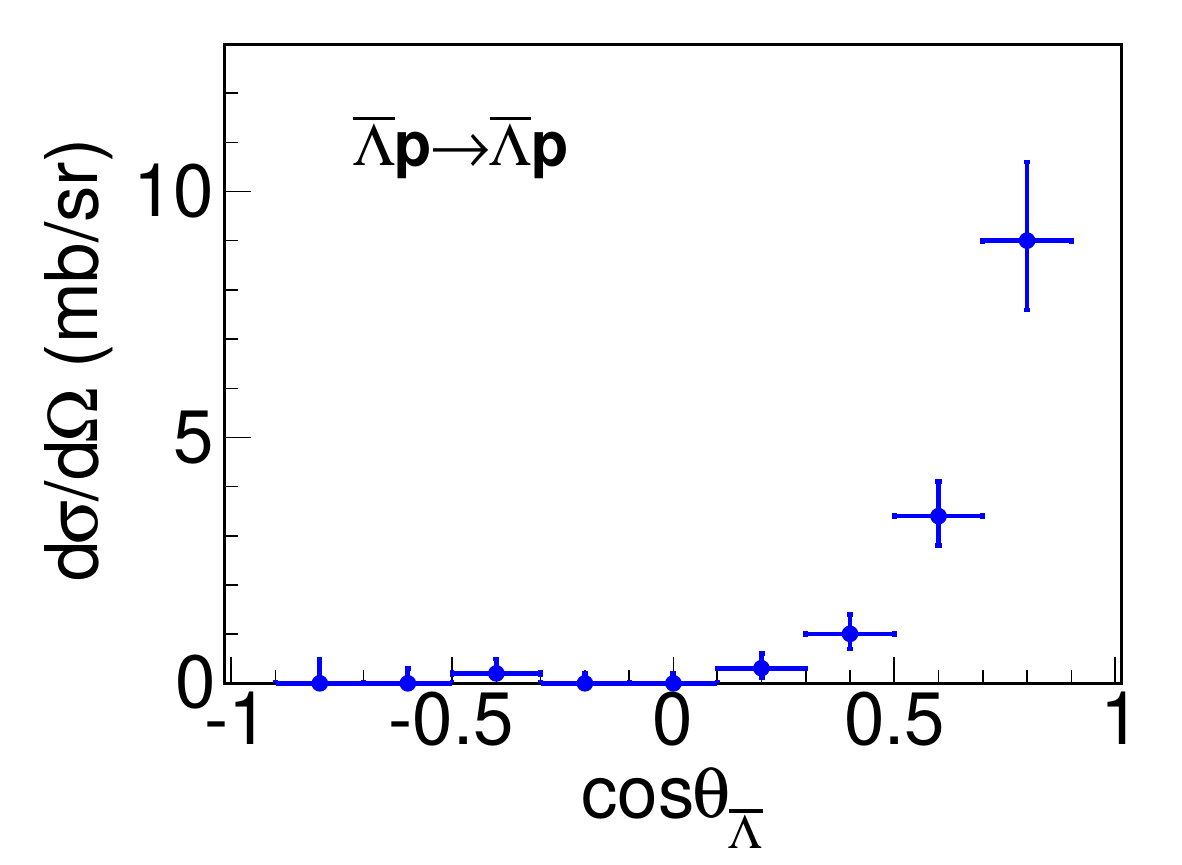}
\end{overpic}
\caption{Differential cross sections of the reactions $\Lambda p\too\Lambda p$ (top) and $\bar{\Lambda}p\too\bar{\Lambda}p$ (bottom) at the $\Lambda/\bar{\Lambda}$ momentum of around $1.074$ GeV/$c$.}
\label{fig:fig_crosssection}
\end{center}
\end{figure}

The differential cross sections for the reactions $\Lambda p\too\Lambda p$ and $\bar{\Lambda}p\too\bar{\Lambda}p$ are calculated with
\begin{equation}
    (\frac{d\sigma}{d\Omega})_i = \frac{N^{\rm{sig}}_{i}} {\epsilon_{i} \mathcal{B} \mathcal{L}_{\rm{eff}} \Delta\Omega},
\end{equation}
where $N^{\rm{sig}}_i$ and $\epsilon_i$ are the number of signal events and efficiency, $i$ represents different cos$\theta_{\Lambda/\bar{\Lambda}}$ bins, and $\Delta\Omega=2\pi\Delta\rm{cos}\theta_{\Lambda/\bar{\Lambda}}=0.4\pi$ represents the solid angle. The measured results are listed in Table~\ref{tab:result} and shown in Fig.~\ref{fig:fig_crosssection}. We can see there is a slight tendency of forward scattering for $\Lambda p\too\Lambda p$, while there is a strong forward peak for $\bar{\Lambda}p\too\bar{\Lambda}p$. The different behaviors indicate that the reaction mechanisms of these two processes are different.

We also tested an extrapolation for the regions of $|\rm{cos}\theta_{\Lambda/\bar{\Lambda}}|>0.9$ for the differential cross sections of $\Lambda p\too\Lambda p$ and $\bar{\Lambda} p\too\bar{\Lambda} p$ to determine the total elastic cross sections. For the reaction $\Lambda p\too\Lambda p$, we assume the differential cross sections in $-1.0<\rm{cos}\theta_{\Lambda}<-0.9$ and $0.9<\rm{cos}\theta_{\Lambda}<1.0$ to be the same as those in neighbouring bins. For the reaction $\bar{\Lambda}p\too\bar{\Lambda}p$, the differential cross section is fitted using a piecewise polynomial function, which is a constant for $\rm{cos}\theta_{\bar{\Lambda}}\leq0$ and a third-order polynomial function for $\rm{cos}\theta_{\bar{\Lambda}}\geq0$. The differential cross section in the regions of $|\rm{cos}\theta_{\bar{\Lambda}}|>0.9$ is obtained according to the fit function. Therefore, the total elastic cross sections integrated over the full angular region are determined to be $\sigma_t(\Lambda p\too\Lambda p)=(14.2\pm1.8_{\rm{stat}}\pm1.3_{\rm{sys}})$~mb and $\sigma_t(\bar{\Lambda} p\too\bar{\Lambda} p)=(27.4\pm3.2_{\rm{stat}}\pm2.5_{\rm{sys}})$~mb. The result of the total elastic cross section on the reaction $\Lambda p\too\Lambda p$ is consistent with those measured from other experiments~\cite{scattering1, scattering2, scattering3, scattering4, scattering5, scattering6, scattering7, scattering8, scattering9, scattering10, scattering11, scattering12, scattering13, scattering14, clas}. The strong forward rise of the differential cross section of $\bar{\Lambda}p\too\bar{\Lambda}p$ is compatible with the expectation for the case of scattering in the presence of a strong absorption~\cite{blacksphere, blacksphere2, blacksphere3}, which is given by the annihilation part of the potential. Especially, this behavior is very similar to $\bar{p}p$ elastic scattering in a comparable incident momentum region~\cite{blacksphere2}, in contrast, such a strong forward rise does not appear in $pp$ elastic scattering~\cite{blacksphere4}. This indicates that the strong absorption mechanism is not only important in $\rm{\bar{N}}$N scattering, but also in $\rm{\bar{Y}}$N scattering. If we assume the reaction $\bar{\Lambda}p\too\bar{\Lambda}p$ is a pure ``black sphere" scattering, the total elastic cross section is given by $\sigma_t(\bar{\Lambda}p\too\bar{\Lambda}p)=\pi R^2$~\cite{blacksphere}, where $R$ is the interaction radius. This gives $R=(0.93\pm0.07)$~fm, which is comparable to the proton radius~\cite{pdg}.

The sources of systematic uncertainties related to the measured cross sections are discussed in the following. The uncertainties in the tracking efficiency and PID efficiency are $1\%$ per track~\cite{xiscatteringbes}. The uncertainty of the track number requirement is estimated with the control sample $J/\psi\too\Lambda\bar{\Lambda}\too p\pi^-\bar{p}\pi^+$. The uncertainties for the branching fractions are taken from the PDG~\cite{pdg}. To estimate the uncertainty from the position of the $\EE$ interaction point, we change the integral range by $\pm0.1$~cm, which is from $(a, b)$ to $(a+0.1, b+0.1)$ or $(a-0.1, b-0.1)$, and the larger difference in the result is taken as the uncertainty. The systematic uncertainties from $\Lambda/\bar{\Lambda}$ mass windows, $M_{\text{recoil}}(\bar{\Lambda}p_{\Lambda})/M_{\text{recoil}}(\Lambda\bar{p})$ requirement, $R_{xy}$ requirement and $P(p_{\text{oil}})$ requirement are tested using a Barlow test method~\cite{lambdascatteringbes}, and these items can be considered negligible. The systematic uncertainties from the fit procedure, the number of $J/\psi$ events, the angular distribution of $J/\psi\too\Lambda\bar{\Lambda}$ and the $\Lambda$ mean lifetime are all less than $1\%$ and can be ignored. A summary of the main systematic uncertainties is presented in Table~\ref{tab:sumerror}, and the total systematic uncertainty is obtained by adding all the individual components in quadrature.
\begin{table}[htbp]
\begin{center}
\caption{Summary of systematic uncertainties (in $\%$).}
\label{tab:sumerror}
\begin{tabular}{cc}
  \hline
  \hline

   Source                     & $\sigma(\Lambda p\too\Lambda p/\bar{\Lambda}p\too\bar{\Lambda}p)$ \\
  \hline
  Tracking efficiency        & 5.0 \\
  PID efficiency             & 5.0 \\
  Track number               & 2.2 \\
  Branching fractions        & 4.9 \\
  $e^+e^-$ interaction point & 2.0 \\
  \hline
  sum                        & 9.1 \\
  \hline
  \hline
\end{tabular}
\end{center}
\end{table}

In summary, using $(10.087\pm0.044)\times10^{9}$ $J/\psi$ events collected with the BESIII detector operating at the BEPCII storage ring, the reactions $\Lambda p\too\Lambda p$ and $\bar{\Lambda}p\too\bar{\Lambda}p$ are measured, where $\Lambda/\bar{\Lambda}$ are from the process $J/\psi\too\Lambda\bar{\Lambda}$ and $p$ is from the cooling oil in the beam pipe. The cross sections in $-0.9\leq\rm{cos}\theta_{\Lambda/\bar{\Lambda}}\leq0.9$ are measured to be $\sigma(\Lambda p\too\Lambda p)=(12.2\pm1.6_{\rm{stat}}\pm1.1_{\rm{sys}})$~mb and $\sigma(\bar{\Lambda} p\too\bar{\Lambda} p)=(17.5\pm2.1_{\rm{stat}}\pm1.6_{\rm{sys}})$~mb at the $\Lambda/\bar{\Lambda}$ momentum of $1.074$~GeV/$c$ within a range of $\pm0.017$~GeV/$c$. Furthermore, the differential cross sections of the two reactions are also measured. There is a slight tendency of forward scattering for $\Lambda p\too\Lambda p$, while a strong forward peak for $\bar{\Lambda}p\too\bar{\Lambda}p$ is observed. If we make an extrapolation for the regions of $|\rm{cos}\theta_{\Lambda/\bar{\Lambda}}|>0.9$ for the differential cross sections of $\Lambda p\too\Lambda p$ and $\bar{\Lambda} p\too\bar{\Lambda} p$, the total elastic cross sections integrated over the full angular region are determined to be $\sigma_t(\Lambda p\too\Lambda p)=(14.2\pm1.8_{\rm{stat}}\pm1.3_{\rm{sys}})$~mb and $\sigma_t(\bar{\Lambda} p\too\bar{\Lambda} p)=(27.4\pm3.2_{\rm{stat}}\pm2.5_{\rm{sys}})$~mb. These constitute the first result of $\rm{\bar{Y}}$N scattering, and will serve as input for the theoretical understanding of the (anti)hyperon-nucleon interaction. This work is the first study of (anti)hyperon-nucleon elastic scattering at an electron-positron collider, and demonstrates the feasibility for studying other antihyperons, such as $\bar{\Sigma} p\too\bar{\Sigma} p$ and $\bar{\Xi} p\too\bar{\Xi} p$. The momentum dependence of these cross sections could be studied at a future super tau-charm factory~\cite{super1, super2} by exploiting multi-body processes or other charmonium decays.

The BESIII Collaboration thanks the staff of BEPCII and the IHEP computing center for their strong support. This work is supported in part by National Key R\&D Program of China under Contracts Nos. 2020YFA0406300, 2020YFA0406400; National Natural Science Foundation of China (NSFC) under Contracts Nos. 12375071, 11635010, 11735014, 11835012, 11935015, 11935016, 11935018, 11961141012, 12025502, 12035009, 12035013, 12061131003, 12192260, 12192261, 12192262, 12192263, 12192264, 12192265, 12221005, 12225509, 12235017; Natural Science Foundation of Henan under Contract No. 242300421163; the Chinese Academy of Sciences (CAS) Large-Scale Scientific Facility Program; the CAS Center for Excellence in Particle Physics (CCEPP); Joint Large-Scale Scientific Facility Funds of the NSFC and CAS under Contract No. U1832207; CAS Key Research Program of Frontier Sciences under Contracts Nos. QYZDJ-SSW-SLH003, QYZDJ-SSW-SLH040; 100 Talents Program of CAS; The Institute of Nuclear and Particle Physics (INPAC) and Shanghai Key Laboratory for Particle Physics and Cosmology; European Union's Horizon 2020 research and innovation programme under Marie Sklodowska-Curie grant agreement under Contract No. 894790; German Research Foundation DFG under Contracts Nos. 455635585, Collaborative Research Center CRC 1044, FOR5327, GRK 2149; Istituto Nazionale di Fisica Nucleare, Italy; Ministry of Development of Turkey under Contract No. DPT2006K-120470; National Research Foundation of Korea under Contract No. NRF-2022R1A2C1092335; National Science and Technology fund of Mongolia; National Science Research and Innovation Fund (NSRF) via the Program Management Unit for Human Resources \& Institutional Development, Research and Innovation of Thailand under Contract No. B16F640076; Polish National Science Centre under Contract No. 2019/35/O/ST2/02907; The Swedish Research Council; U. S. Department of Energy under Contract No. DE-FG02-05ER41374.

\end{document}